\documentclass[11pt,preprintnumbers,titlepage,nofootinbib,showkeys,showpacs]{revtex4-2}
\usepackage[latin9]{inputenc}
\usepackage{amsmath}
\usepackage{amssymb}
\usepackage{graphicx}
\usepackage[unicode=true,
 bookmarks=false,
 breaklinks=false,pdfborder={0 0 1},backref=section,colorlinks=false]
 {hyperref}
\hypersetup{
 colorlinks,linkcolor=blue,citecolor=blue,urlcolor=blue}

\makeatletter
\usepackage{wasysym}
\usepackage{amsfonts}
\usepackage{subfigure}
\usepackage{cleveref}
\usepackage{listings}
\usepackage{xcolor}
\usepackage{array}
\usepackage{url}
\@ifundefined{textcolor}{}{
\definecolor{BLACK}{gray}{0}
\definecolor{WHITE}{gray}{1}
\definecolor{RED}{rgb}{1,0,0}
\definecolor{GREEN}{rgb}{0,1,0}
\definecolor{BLUE}{rgb}{0,0,1}
\definecolor{CYAN}{cmyk}{1,0,0,0}
\definecolor{MAGENTA}{cmyk}{0,1,0,0}
\definecolor{YELLOW}{cmyk}{0,0,1,0}
}

\makeatother

\begin{document}
\preprint{CTP-SCU/2023040}
\title{Interferometric Signatures of Black Holes with Multiple Photon Spheres}
\author{Yiqian Chen}
\email{chenyiqian@stu.scu.edu.cn}

\author{Peng Wang}
\email{pengw@scu.edu.cn}

\author{Haitang Yang}
\email{hyanga@scu.edu.cn}

\affiliation{Center for Theoretical Physics, College of Physics, Sichuan University,
Chengdu, 610064, China}
\begin{abstract}
It has been reported that the photon ring structure in black hole
images produces strong and universal interferometric signatures on
long interferometric baselines, holding promise for measuring black
hole parameters and testing general relativity. This paper investigates
the interferometric signatures of black holes with one or two photon
spheres, specifically within the framework of Einstein-Maxwell-Scalar
models. Notably, for black holes possessing two photon spheres, interference
between light rays orbiting the inner and outer photon spheres manifests
as beat signals in the visibility amplitude, deviating from the universal
signatures observed in the single-photon sphere case.
\end{abstract}
\maketitle
\tableofcontents{}

\section{Introduction}

\label{sec:Introduction}

Recently, the Event Horizon Telescope (EHT) collaboration has released
images of the supermassive black holes M87{*} \cite{Akiyama:2019cqa,Akiyama:2019brx,Akiyama:2019sww,Akiyama:2019bqs,Akiyama:2019fyp,Akiyama:2019eap,Akiyama:2021qum,Akiyama:2021tfw}
and Sgr A{*} \cite{EventHorizonTelescope:2022xnr,EventHorizonTelescope:2022vjs,EventHorizonTelescope:2022wok,EventHorizonTelescope:2022exc,EventHorizonTelescope:2022urf,EventHorizonTelescope:2022xqj},
both of which exhibit a bright ring encircling a dark shadow. These
prominent features are closely connected to strong light deflections
that occur in proximity to unstable bound photon orbits, which form
photon spheres in spherically symmetric scenarios \cite{Synge:1966okc,Bardeen:1972fi,Bardeen:1973tla,Bozza:2009yw}.
While the EHT has successfully measured the diameter of the bright
ring, its detailed substructure remains unresolved. There is a strong
belief that this substructure holds valuable information about the
geometry near black holes, particularly in the vicinity of the event
horizon.

Disregarding opacity, photons orbiting the black hole multiple times
generate an infinite series of nested images of the accretion disk.
These images asymptotically approach the critical curves, formed by
light rays escaping from bound photon orbits. In \cite{Gralla:2019xty},
emission from an optically and geometrically thin disk is categorized
into three classes: direct emission (primary image), lensing ring
(secondary image) and photon ring (tertiary and higher-order images).
Notably, the photon ring comprises an infinite sequence of self-similar
subrings. Their analysis demonstrates that the superposition of the
lensing and photon rings on the direct emission produces a bright
ring in the black hole image. Furthermore, the photon ring's structure
exhibits remarkable insensitivity to the astronomical source, making
it a valuable tool for measuring black hole parameters and testing
general relativity \cite{Gralla:2020srx,Wielgus:2021peu,Li:2021mzq,Broderick:2021ohx}.
However, the limited resolution of the EHT poses a challenge in resolving
the photon ring within black hole images due to its exponentially
narrow profile.

Recently, the authors of \cite{Johnson:2019ljv} demonstrated the
feasibility of precise measurements of the photon ring and its subrings
through the analysis of interferometric signatures in black hole images.
The study revealed that the lensing ring of both M87{*} and Sgr A{*}
could be precisely measured using high-frequency ground arrays or
low Earth orbit stations, while the first and second photon subrings
could potentially be measured with lunar and second Sun-Earth Lagrange
point stations, respectively. Furthermore, the photon ring has been
shown to exhibit universal signatures on the complex visibility signal,
with its subrings manifesting as a cascade of damped oscillations
on long baselines. These properties enable the decoding of black hole
metrics from the interferometric pattern of the photon ring via analysis
of the amplitude, width, and period of the complex visibility \cite{Aratore:2021usi}.
Further research in this domain has explored various aspects of the
photon ring, including the autocorrelations of self-similar photon
rings \cite{Hadar:2020fda,Chesler:2020gtw,Qian:2021aju}, polarimetric
signatures \cite{Himwich:2020msm,Gussmann:2021mjj} and the impact
of different accretion models \cite{Paugnat:2022qzy,Vincent:2022fwj}.

Meanwhile, a class of Einstein-Maxwell-scalar (EMS) models has been
proposed to explain the formation of scalarized black holes \cite{Herdeiro:2018wub}.
These models incorporate non-minimal couplings between the scalar
and Maxwell fields, inducing tachyonic instabilities capable of initiating
spontaneous scalarization. Utilizing fully non-linear numerical simulations,
Herdeiro et al. demonstrated the evolution of Reissner-Nordström (RN)
black holes into scalarized RN black holes \cite{Herdeiro:2018wub}.
This discovery has sparked a wealth of research within the EMS framework,
exploring diverse areas such as different non-minimal coupling functions
\cite{Fernandes:2019rez,Fernandes:2019kmh,Blazquez-Salcedo:2020nhs},
massive and self-interacting scalar fields \cite{Zou:2019bpt,Fernandes:2020gay},
horizonless reflecting stars \cite{Peng:2019cmm}, stability analysis
of scalarized black holes \cite{Myung:2018vug,Myung:2019oua,Zou:2020zxq,Myung:2020etf,Mai:2020sac},
higher dimensional scalar-tensor models \cite{Astefanesei:2020qxk},
quasinormal modes of scalarized black holes \cite{Myung:2018jvi,Blazquez-Salcedo:2020jee},
two U(1) fields \cite{Myung:2020dqt}, quasitopological electromagnetism
\cite{Myung:2020ctt}, topology and spacetime structure influences
\cite{Guo:2020zqm}, scalarized black hole solutions in the dS/AdS
spacetime \cite{Brihaye:2019dck,Brihaye:2019gla,Zhang:2021etr,Guo:2021zed,Chen:2023eru},
dynamical scalarization and descalarization \cite{Zhang:2021nnn,Zhang:2022cmu,Jiang:2023yyn}
and rotating scalarized black hole solutions \cite{Guo:2023mda}.

Remarkably, scalarized RN black holes have been found to exhibit multiple
photon spheres outside the event horizon in specific parameter regimes
\cite{Gan:2021pwu}. Subsequent research has focused on the optical
appearances of various phenomena in this background, including accretion
disks \cite{Gan:2021pwu,Gan:2021xdl}, luminous celestial spheres
\cite{Guo:2022muy} and infalling stars \cite{Chen:2022qrw}. These
investigations have unveiled that the existence of an additional photon
sphere significantly amplifies the flux of observed accretion disk
images, generates triple higher-order images of a luminous celestial
sphere and instigates an additional cascade of flashes from an infalling
star. Furthermore, multiple photon spheres suggest the existence of
long-lived modes potentially rendering the spacetime unstable \cite{Cardoso:2014sna,Keir:2014oka,Guo:2021bcw,Guo:2021enm,Guo:2022umh}.
Specifically, recent work has shown that their presence can induce
superradiance instabilities for charged scalar perturbations \cite{Guo:2023ivz}.
Furthermore, the existence of two photon spheres outside the event
horizon has also been reported in dyonic black holes with a quasi-topological
electromagnetic term \cite{Liu:2019rib,Huang:2021qwe}, black holes
in massive gravity \cite{deRham:2010kj,Dong:2020odp} and wormholes
in the black-bounce spacetime \cite{Tsukamoto:2021caq,Tsukamoto:2021fpp,Tsukamoto:2022vkt}.
For a more detailed analysis of black holes with multiple photon spheres,
we refer readers to \cite{Guo:2022ghl}.

This paper investigates the interferometric signatures of scalarized
RN black hole images, with a specific focus on understanding how an
additional photon sphere influences these signatures. The remainder
of the paper is structured as follows. Section \ref{sec:SRNBH} provides
a succinct overview of scalarized RN black hole solutions within the
EMS model. In Section \ref{sec:BHICV}, we present the method for
obtaining observed images of accretion disks and their corresponding
complex visibility. Subsequently, we discuss the visibility signals
of black holes featuring either one or two photon spheres using simplified
thin-ring models. Section \ref{sec:NS} displays numerical simulations
of accretion disk images in scalarized RN black holes, along with
the corresponding visibility. Section \ref{sec:Conclusions} is devoted
to our main conclusions. We set $G=c=1$ throughout the paper.

\section{Scalarized RN Black Holes}

\label{sec:SRNBH}

In this section, we provide a concise review of the scalarized RN
black hole solution in the 4-dimensional EMS model. This model is
described by the action \cite{Herdeiro:2018wub} 
\begin{equation}
S=\frac{1}{16\pi}\int d^{4}x\sqrt{-g}\left[\mathcal{R}-2\partial_{\mu}\phi\partial^{\mu}\phi-f\left(\phi\right)F_{\mu\nu}F^{\mu\nu}\right],\label{eq:Action}
\end{equation}
where $\mathcal{R}$ is the Ricci scalar, and $F_{\mu\nu}=\partial_{\mu}A_{\nu}-\partial_{\nu}A_{\mu}$
is the electromagnetic field strength tensor. In this EMS model, the
scalar field $\phi$ is non-minimally coupled to the electromagnetic
field $A_{\mu}$ through the coupling function $f\left(\phi\right)$.
To admit scalar-free black hole solutions, specifically RN black holes,
the coupling function must satisfy the condition $f^{\prime}\left(0\right)\equiv\left.df\left(\phi\right)/d\phi\right\vert _{\phi=0}=0$
\cite{Herdeiro:2018wub,Fernandes:2019rez}. In this paper, we focus
on the exponential coupling function $f\left(\phi\right)=e^{\alpha\phi^{2}}$
with $\alpha>0$. Within the RN black hole background, the equation
of motion governing scalar perturbations $\delta\phi$ takes the form,
\begin{equation}
\left(\square-\mu_{\text{eff}}^{2}\right)\delta\phi=0,\label{eq:delta phi}
\end{equation}
where $\mu_{\text{eff}}^{2}=-\alpha Q^{2}/r^{4}$. It is worth noting
that tachyonic instabilities arise when the effective mass square
$\mu_{\text{eff}}^{2}$ becomes negative. These instabilities, as
demonstrated in \cite{Herdeiro:2018wub,Guo:2021zed}, can be sufficiently
strong near the event horizon to induce the formation of scalarized
RN black holes from their scalar-free counterparts.

To derive scalarized RN black holes, we adopt the following generic
ansatz, 
\begin{align}
ds^{2} & =-N\left(r\right)e^{-2\delta\left(r\right)}dt^{2}+\frac{1}{N\left(r\right)}dr^{2}+r^{2}\left(d\theta^{2}+\sin^{2}\theta d\varphi^{2}\right),\nonumber \\
A_{\mu}dx^{\mu} & =\Phi\left(r\right)dt\text{ and}\ \phi=\phi\left(r\right).\label{eq:HBH}
\end{align}
The corresponding equations of motion are given by 
\begin{align}
N^{\prime}\left(r\right) & =\frac{1-N\left(r\right)}{r}-\frac{Q^{2}}{r^{3}e^{\alpha\phi^{2}\left(r\right)}}-rN\left(r\right)\left[\phi^{\prime}\left(r\right)\right]^{2},\nonumber \\
\left[r^{2}N\left(r\right)\phi^{\prime}\left(r\right)\right]^{\prime} & =-\frac{\alpha Q^{2}\phi\left(r\right)}{r^{2}e^{\alpha\phi^{2}\left(r\right)}}-r^{3}N\left(r\right)\left[\phi^{\prime}\left(r\right)\right]^{3},\nonumber \\
\delta^{\prime}\left(r\right) & =-r\left[\phi^{\prime}\left(r\right)\right]^{2},\label{eq:EOM}\\
\Phi^{\prime}\left(r\right) & =\frac{Q}{r^{2}e^{\alpha\phi^{2}\left(r\right)}}e^{-\delta\left(r\right)},\nonumber 
\end{align}
where the integration constant $Q$ can be interpreted as the electric
charge of the black hole, and primes stand for derivative with respect
to $r$. Proper boundary conditions are imposed at the event horizon
$r_{h}$ and spatial infinity as 
\begin{align}
N\left(r_{h}\right) & =0\text{, }\delta\left(r_{h}\right)=\delta_{0}\text{, }\phi\left(r_{h}\right)=\phi_{0}\text{, }\Phi\left(r_{h}\right)=0\text{,}\nonumber \\
N\left(\infty\right) & =1-\frac{2M}{r}\text{, }\delta\left(\infty\right)=0\text{, }\phi\left(\infty\right)=0\text{, }\Phi\left(\infty\right)=\Psi\text{,}\label{eq:infinity condition}
\end{align}
where $\delta_{0}$ and $\phi_{0}$ can be used to characterize black
hole solutions, $M$ is the ADM mass of the black hole, and $\Psi$
is the electrostatic potential. After specifying $\delta_{0}$ and
$\phi_{0}$, scalarized RN black hole solutions with a non-trivial
scalar field $\phi$ are obtained using the shooting method incorporated
in the $NDSolve$ function of $Wolfram\text{ }\circledR Mathematica$.
For simplicity and without loss of generality, we express all physical
quantities in units of the black hole mass by setting $M=1$ throughout
the paper.

The trajectory of a light ray in scalarized RN black holes follows
the null geodesic equation, 
\begin{equation}
\frac{d^{2}x^{\mu}}{d\tau^{2}}+\Gamma_{\rho\sigma}^{\mu}\frac{dx^{\rho}}{d\tau}\frac{dx^{\sigma}}{d\tau}=0,\label{eq:geo eq}
\end{equation}
where $\tau$ is the affine parameter, and $\Gamma_{\rho\sigma}^{\mu}$
represents the Christoffel symbol. Given the spherical symmetry, our
analysis focuses on the equatorial plane. Combining the null geodesic
condition $ds^{2}=0$ with eqn. $\left(\ref{eq:geo eq}\right)$, the
radial components of the null geodesic equation can be derived as
\begin{equation}
\frac{e^{-2\delta\left(r\right)}}{L^{2}}\left(\frac{dr}{d\lambda}\right)^{2}+V_{\text{eff}}\left(r\right)=\frac{1}{b^{2}},
\end{equation}
where $V_{\text{eff}}\equiv e^{-2\delta\left(r\right)}N\left(r\right)r^{-2}$
is the effective potential, $b\equiv\left\vert L\right\vert /E$ is
the impact parameter, and $L$ and $E$ are the conserved angular
momentum and energy of photons, respectively. In spherical symmetric
black holes, unstable circular null geodesics constitute a photon
sphere of radius $r_{\text{ph}}$, determined by 
\begin{equation}
V_{\text{eff}}\left(r_{\text{ph}}\right)=b_{\text{ph}}^{-2}\text{, }V^{\prime}\left(r_{\text{ph}}\right)=0\text{, }V^{\prime\prime}\left(r_{\text{ph}}\right)<0,
\end{equation}
where $b_{\text{ph}}$ is the corresponding impact parameter. In summary,
a local maximum of the effective potential signifies the presence
of a photon sphere.

\begin{figure}[ptb]
\includegraphics[width=0.45\textwidth]{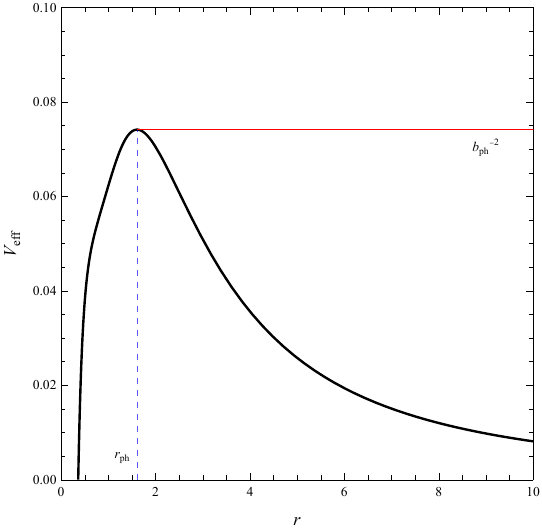} \hspace{20pt}
\includegraphics[width=0.45\textwidth]{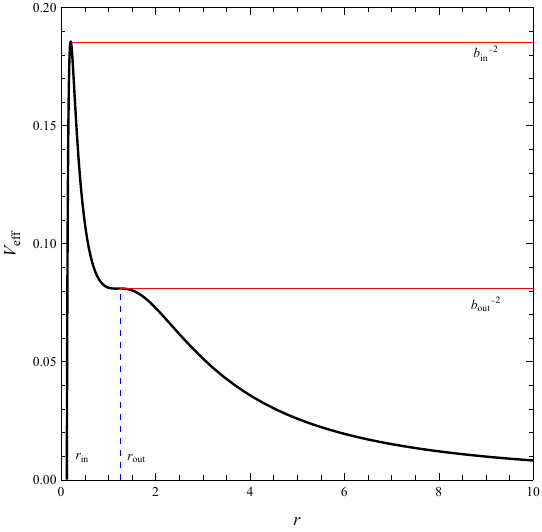} \caption{\textbf{Left:} The effective potential of the scalarized RN black
hole with $\alpha=0.9$ and $Q=1.054$. A single potential maximum
is observed at $r_{\text{ph}}=1.610$, corresponding to the photon
sphere with a critical impact parameter $b_{\text{ph}}=3.673$. \textbf{Right:}
The effective potential of the scalarized RN black hole with $\alpha=0.9$
and $Q=1.070$, displaying two maxima. The potential maxima represent
two photon spheres, situated at $r_{\text{in}}=0.204$ and $r_{\text{out}}=1.255$
with critical impact parameters $b_{\text{in}}=2.322$ and $b_{\text{out}}=3.515$,
respectively. The considerable height difference between these two
peaks indicates a relatively large separation between the two critical
curves on the black hole image plane.}
\label{fig:veff}
\end{figure}

Remarkably, our previous study revealed that scalarized RN black holes
can possess either one or two photon spheres outside the event horizon,
contingent on the values of $\alpha$ and $Q/M$ \cite{Gan:2021pwu}.
The left panel of FIG. \ref{fig:veff} illustrates the effective potential
of a scalarized RN black hole with $\alpha=0.9$ and $Q=1.054$, featuring
a single peak. Conversely, the right panel of FIG. \ref{fig:veff}
depicts the effective potential of a scalarized RN black hole with
$\alpha=0.9$ and $Q=1.070$, manifesting a double-peak structure.
In our previous work \cite{Guo:2022ghl}, we delineated the region
in the $\alpha$-$Q/M$ parameter space where double photon spheres
exist.

\section{Black Hole Image and Complex Visibility}

\label{sec:BHICV}

In this paper, we investigate scalarized RN black holes encompassed
by a static and geometrically thin accretion disk, presumed to radiate
isotropically in the rest frame of the matter. Additionally, we consider
the accretion disk emission to be optically thin, disregarding absorption
effects. This assumption is supported by compelling evidence suggesting
the presence of an optically thin hot accretion flow around black
holes such as M87{*} or Sgr A{*} \cite{Yuan:2014gma,Johnson:2015iwg}.
To obtain the image of the accretion disk observed by a distant observer,
light rays are evolved backwards in time from the observer's position.
Assuming the observer is located on the disk's axis of symmetry, the
observed intensity $I_{\text{o}}$ exhibits axial symmetry on the
image plane defined by the observation angles. By setting the origin
of the image plane at the black hole center, we introduce polar coordinates
$\mathbf{x}=\left(\beta,\phi_{\beta}\right)$, where $\beta$ represents
the angle between the incident ray and the disk's axis of symmetry.
Consequently, the observed intensity $I_{\text{o}}$ becomes a function
solely of $\beta$.

The observed intensity $I_{\text{o}}\left(\beta\right)$, resulting
from light rays with an incident angle $\beta$, is a summation of
intensities acquired at each intersection with the disk plane beyond
the horizon, 
\begin{equation}
I_{\text{o}}\left(\beta\right)=\underset{n}{\sum}g^{4}\left(r_{n}\left(\beta\right),\beta\right)I_{\text{em}}\left(r_{n}\left(\beta\right)\right).\label{eq:Observed I}
\end{equation}
Here, $g$ represents the redshift factor between the observed frequency
and emission frequency \cite{Gralla:2020srx}, $I_{\text{em}}\left(r\right)$
stands for the emitted intensity at radial coordinate $r$, and $r_{n}\left(\beta\right)$
denotes the transfer function representing the radial coordinate of
the $n^{\text{th}}$ intersection with the thin accretion disk. The
index $n$ corresponds to the number of half-orbits, defined as $\Delta\varphi/\pi+1/2$,
where $\Delta\varphi$ represents the angular variation of light rays.
Following the terminology in \cite{Gralla:2019xty}, light rays with
$n=1$, $n=2$ and $n\geq3$ constitute the direct emission, lensing
ring and photon ring of black hole images, respectively. It is noteworthy
that the photon ring comprises an infinite series of subrings with
a specific $n$. As $n$ increases, the photon subring converges toward
a circular critical curve formed by light rays escaping from a photon
sphere.

If an interferometer measures the black hole image, each baseline
joining two elements of the interferometer samples a single Fourier
component of the observed intensity $I_{\text{o}}(\mathbf{x})$, referred
to as the complex visibility \cite{Thompson2017}, 
\begin{equation}
V\left(\mathbf{u}\right)=\int I_{\text{o}}\left(\mathbf{x}\right)e^{-2\pi i\mathbf{u}\cdot\mathbf{x}}d^{2}\mathbf{x},\label{eq:visibility}
\end{equation}
where $\mathbf{u}$ represents the dimensionless baseline vector projected
orthogonally to the line of sight and measured in units of the observation
wavelength. In the case of the observed intensity $\left(\ref{eq:Observed I}\right)$,
the complex visibility simplifies to a zero-order Hankel transform,
\begin{equation}
V\left(u\right)=2\pi\int I_{\text{o}}\left(\beta\right)J_{0}\left(2\pi\beta u\right)\beta d\beta,\label{eq:HT}
\end{equation}
where $u=\left\vert \mathbf{u}\right\vert $ denotes the baseline
length, and $J_{0}$ represents the zero-order Bessel function of
the first kind.

In the scenario where a single photon sphere exists, light rays originating
from the photon sphere collectively give rise to a circular critical
curve on the image plane, with a radius denoted as $\beta_{\text{ph}}$.
As demonstrated in \cite{Gralla:2019xty}, the photon ring closely
follows this critical curve, with a width significantly smaller than
those of the direct emission and lensing ring. Consequently, we initially
adopt a simplified model, treating the photon ring as a thin, uniform
and circular ring with a radius $\beta_{\text{ph}}$ on the image
plane. Specifically, the intensity of the photon ring $I_{\text{o}}^{\text{pr}}\left(\beta\right)$
can be expressed as
\begin{equation}
I_{\text{o}}^{\text{pr}}\left(\beta\right)=\Phi_{\text{ph}}\delta\left(\beta-\beta_{\text{ph}}\right),
\end{equation}
where $\Phi_{\text{ph}}$ represents a constant. The corresponding
complex visibility is given by
\begin{equation}
V\left(u\right)=2\pi\Phi_{\text{ph}}\beta_{\text{ph}}J_{0}\left(2\pi\beta_{\text{ph}}u\right).
\end{equation}
For long baselines ($u\gg1/\left(16\pi\beta_{\text{ph}}\right)$),
the complex visibility can be approximated as
\begin{equation}
V\left(u\right)\approx2\Phi_{\text{ph}}\sqrt{\beta_{\text{ph}}}\frac{\cos\left(2\pi\beta_{\text{ph}}u-\pi/4\right)}{\sqrt{u}},
\end{equation}
which describes a weakly damped oscillation with a period $1/\beta_{\text{ph}}$
decaying as $1/\sqrt{u}$. When the internal structure of the photon
ring is considered, the $n^{\text{th}}$ subring of width $w_{n}$
dominates the visibility signal in the regime $1/w_{n-1}\ll u\ll1/w_{n}$,
resulting in a cascade of damped oscillations for the complex visibility
\cite{Johnson:2019ljv}.

In the scenario involving double photon spheres, both photon spheres
can contribute to black hole images when the potential peak at the
inner photon sphere is higher than that at the outer one. In such
cases, the inner and outer photon spheres manifest as an inner critical
curve at $\beta=\beta_{\text{in}}$ and an outer one at $\beta=\beta_{\text{out}}$,
respectively, on black hole images. Furthermore, the intensity of
the photon ring exhibits two sharp peaks at these locations, corresponding
to bound photons orbiting each photon sphere multiple times \cite{Gan:2021xdl}.
Similarly to the single-photon sphere case, the intensity of the photon
ring can be approximated as two infinitesimally thin, uniform and
circular rings,
\begin{equation}
I_{\text{o}}^{\text{pr}}\left(\beta\right)=\Phi_{\text{in}}\delta\left(\beta-\beta_{\text{in}}\right)+\Phi_{\text{out}}\delta\left(\beta-\beta_{\text{out}}\right),
\end{equation}
where $\Phi_{\text{in}}$ and $\Phi_{\text{out}}$ are constants.
Applying eqn. $\left(\ref{eq:HT}\right)$ provides the complex visibility,
\begin{equation}
V\left(u\right)=2\pi\Phi_{\text{in}}\beta_{\text{in}}J_{0}\left(2\pi\beta_{\text{in}}u\right)+2\pi\Phi_{\text{out}}\beta_{\text{out}}J_{0}\left(2\pi\beta_{\text{out}}u\right),
\end{equation}
which, for $u\gg1/\left(16\pi\beta_{\text{ph}}\right)$, becomes
\begin{equation}
V\left(u\right)\approx2\Phi_{\text{in}}\sqrt{\beta_{\text{in}}}\frac{\cos\left(2\pi\beta_{\text{in}}u-\pi/4\right)}{\sqrt{u}}+2\Phi_{\text{out}}\sqrt{\beta_{\text{out}}}\frac{\cos\left(2\pi\beta_{\text{out}}u-\pi/4\right)}{\sqrt{u}}.
\end{equation}
The resulting equation reveals beat signals in the visibility with
a period $\Delta u=1/\left(\beta_{\text{out}}-\beta_{\text{in}}\right)$,
which is subsequently affirmed through numerical simulations of black
hole images.

\section{Numerical Simulations}

\label{sec:NS}

In this section, we conduct numerical simulations to generate images
of scalarized RN black holes exhibiting either a single or two photon
spheres, along with their associated interferometric signatures. An
accretion disk is situated on the equatorial plane and observed by
a distant observer at $r_{\text{o}}=100$ in a face-on orientation.
To streamline the analysis, we adopt a total emitted intensity $I_{\text{em}}(r)=1/r^{2}$
extending to the horizon, providing sufficient clarity to illustrate
the primary features of the accretion disk images.

\begin{figure}[ptb]
\includegraphics[width=0.415\textwidth]{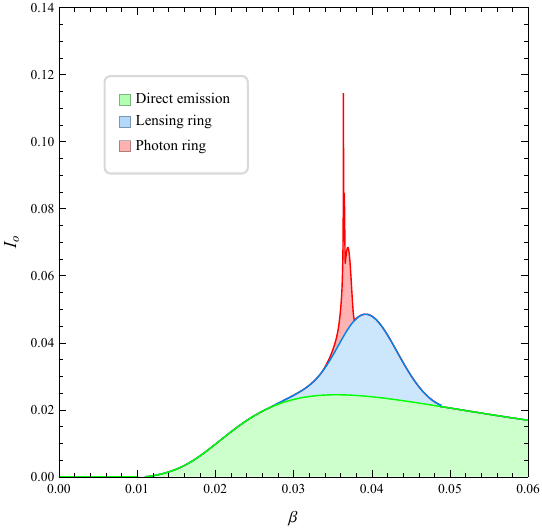} \hspace{8.5pt}
\includegraphics[width=0.4\textwidth]{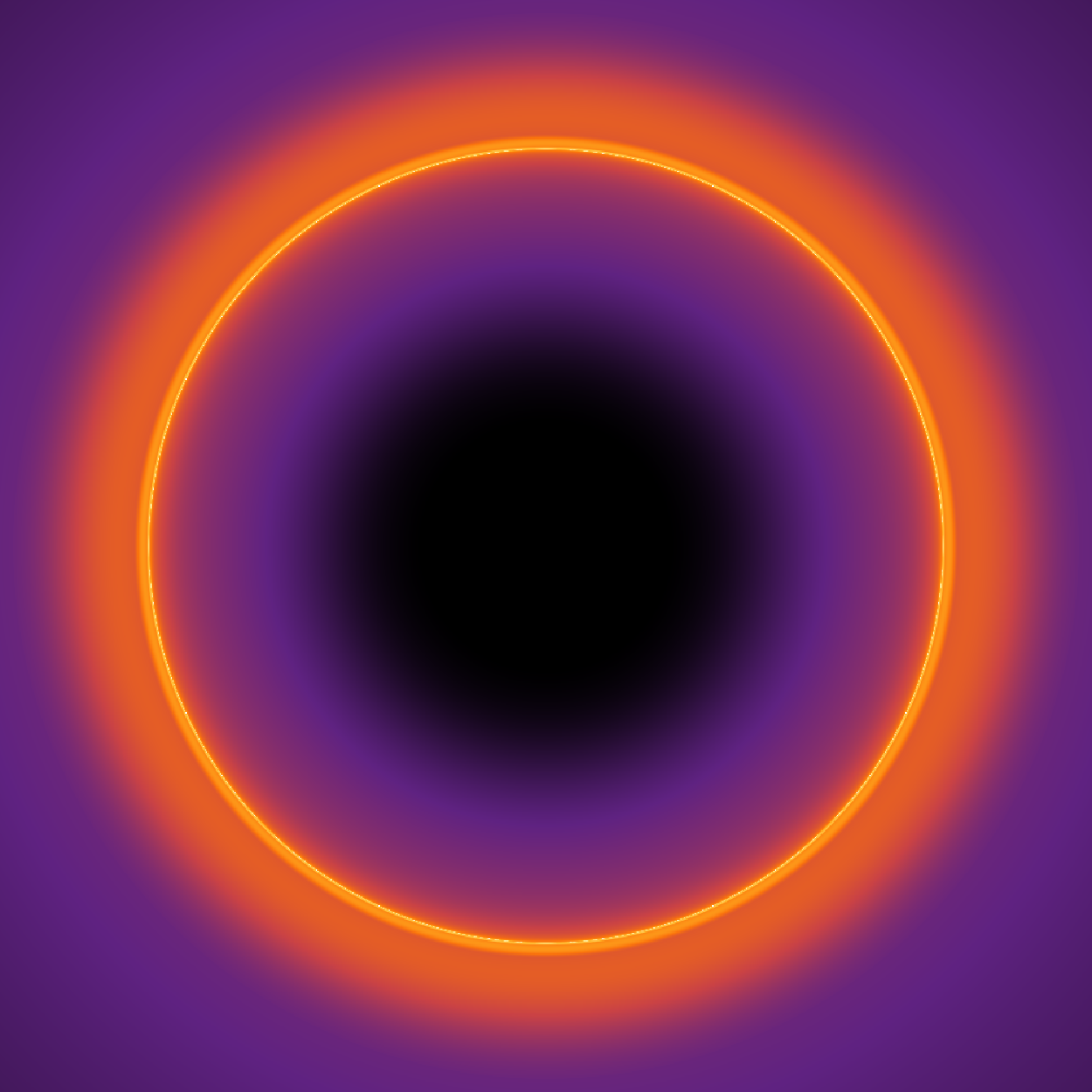} 

\smallskip{}
 \smallskip{}

\includegraphics[width=0.87\textwidth]{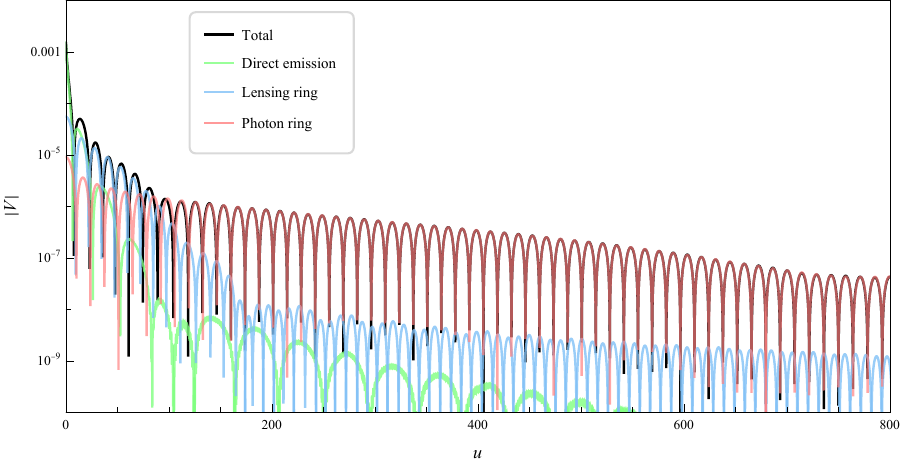}\caption{Scalarized RN black hole with $\alpha=0.9$ and $Q=1.054$, featuring
a single photon sphere. \textbf{Top-left: }Observed intensity $I_{\text{o}}\left(\beta\right)$
of the black hole viewed face-on by a distant observer at $r_{\text{o}}=100$,
decomposed into contributions from the direct emission (green), lensing
ring (blue) and photon ring (red). The inner edge of the intensity
profile corresponds to the lensed position of the event horizon at
$\beta\simeq0.011$, with the central dark region referred to as the
"inner shadow". \textbf{Top-right: }Image of the accretion disk
viewed in the observer's sky. A bright and narrow ring appears close
to the critical curve, primarily arising from the direct emission
and lensing ring. \textbf{Bottom}: Visibility amplitude $\left\vert V\left(u\right)\right\vert $
and its components from the direct emission (green), lensing ring
(blue) and photon ring (red). Short baselines reveal dominance by
the direct emission and lensing ring, while longer baselines are dominated
by damped oscillations originating from the photon ring.}
\label{fig:sp}
\end{figure}

We begin by examining a scalarized RN black hole with $\alpha=0.9$
and $Q=1.054$, displaying a single-peak potential as depicted in
the left panel of FIG. \ref{fig:veff}. The photon sphere, marked
by the potential's maximum, is situated at $r_{\text{ph}}=1.610$,
with the corresponding critical impact parameter $b_{\text{ph}}=3.673$.
The observed intensity $I_{\text{o}}\left(\beta\right)$ is presented
in the top-left panel of FIG. \ref{fig:sp}, where the green, blue
and red regions denote contributions from the direct emission, lensing
ring and photon ring, respectively. Notably, $I_{\text{o}}\left(\beta\right)$
exhibits a narrow spike-like photon ring and a broader bump-like lensing
ring superimposed on the direct emission. The top-right panel of FIG.
\ref{fig:sp} illustrates the corresponding black hole image, featuring
a bright ring in proximity to the critical curve. It's worth noting
that the photon ring contributes minimally to the brightness of the
ring due to significant demagnification.

The bottom panel of FIG. \ref{fig:sp} presents the total visibility
amplitude alongside the individual contributions from the direct emission,
lensing ring and photon ring. As detailed in \cite{Johnson:2019ljv},
sufficiently long baselines ($u\gg1/w$) are capable of resolving
the intensity profile of any smooth ring with width $w$. Consequently,
the visibility of the ring decays exponentially when $u\gg1/w$. This
phenomenon leads to the dominance of individual contributions within
certain baseline ranges. To be precise, the direct emission prevails
in the visibility at short baselines with $u\lesssim10$, the lensing
ring takes over in the range of $30\lesssim u\lesssim70$, and the
photon ring becomes dominant at long baselines with $u\gtrsim100$.
Furthermore, within the displayed range, the contribution of the photon
ring primarily arises from the $n=3$ subring, thus exhibiting a damped
oscillation.

\begin{figure}[ptb]
\includegraphics[width=0.415\textwidth]{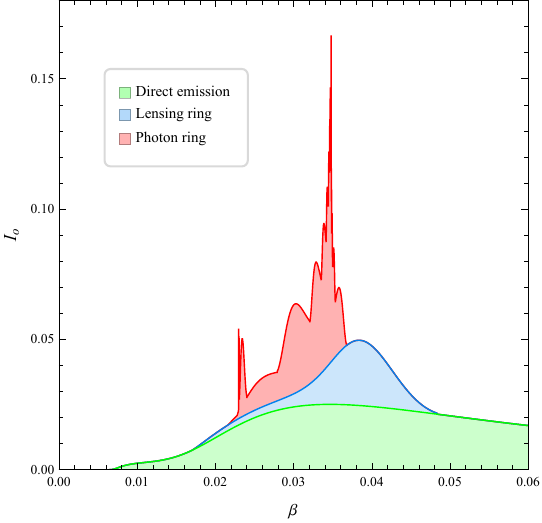} \hspace{8.5pt}
\includegraphics[width=0.4\textwidth]{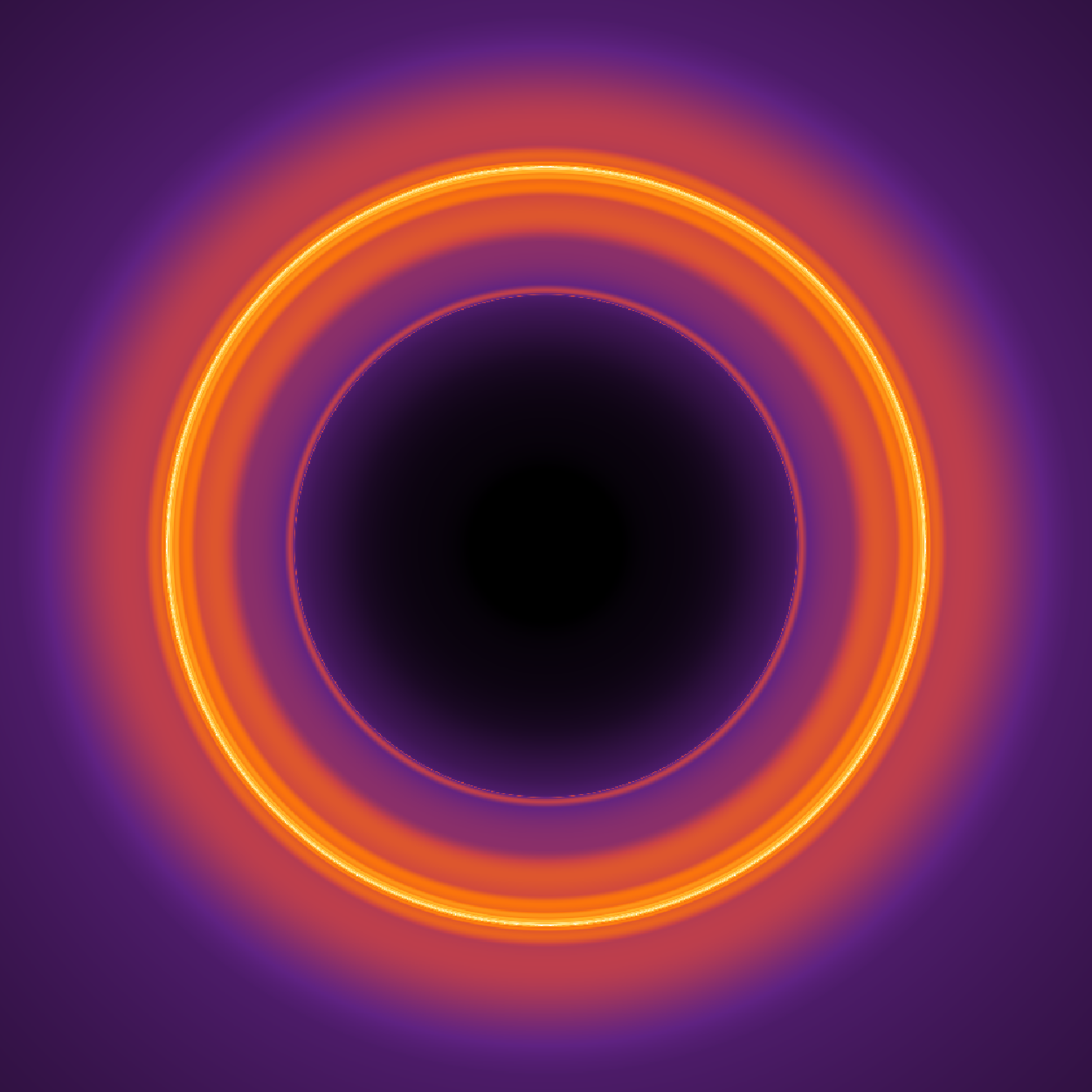} 

\smallskip{}
 \smallskip{}

\includegraphics[width=0.87\textwidth]{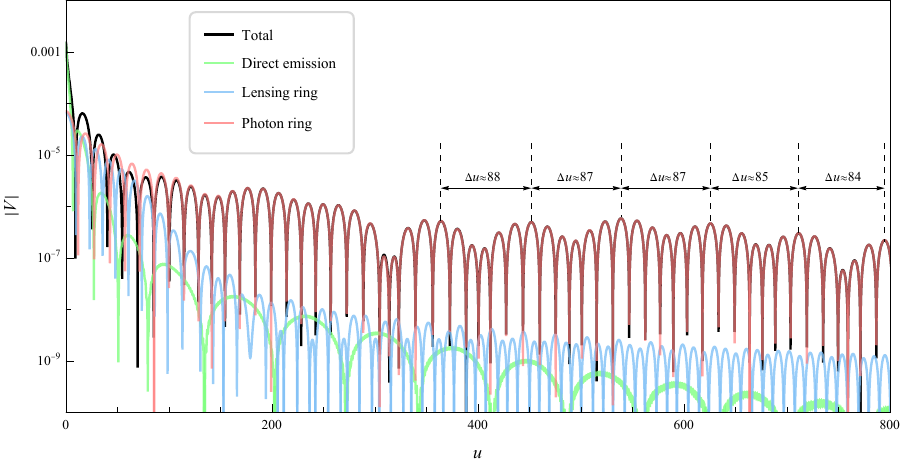}\caption{Scalarized RN black hole with $\alpha=0.9$ and $Q=1.070$, exhibiting
double photon spheres. \textbf{Top-left: }Observed intensity $I_{\text{o}}\left(\beta\right)$
of the black hole, as viewed face-on by a distant observer at $r_{\text{o}}=100$,
partitioned into contributions from the direct emission (green), lensing
ring (blue) and photon ring (red). The lensing and photon rings contribute
comparably to $I_{\text{o}}$, with two distinct intensity peaks at
$\beta_{\text{in}}\simeq0.023$ and $\beta_{\text{out}}\simeq0.035$,
originating from nearly bound photons orbiting around the inner and
outer photon spheres, respectively. \textbf{Top-right: }Image of the
accretion disk as observed in the sky of the observer. Due to light
rays traversing between the two photon spheres, the image displays
intricate patterns. \textbf{Bottom}: Visibility amplitude $\left\vert V\left(u\right)\right\vert $
and its constituents from the direct emission (green), lensing ring
(blue) and photon ring (red). For sufficiently long baselines, two
sharp intensity peaks dominate the visibility amplitude, resulting
in observable beat signals. The time intervals between these beat
signals align well with $1/\left(\beta_{\text{out}}-\beta_{\text{in}}\right)\simeq83.33$.}
\label{fig:dp}
\end{figure}

We further explore a scalarized RN black hole with $\alpha=0.9$ and
$Q=1.070$, characterized by a double-peak potential as illustrated
in the right panel of FIG. \ref{fig:veff}. Two photon spheres are
identified at $r_{\text{in}}=0.204$ and $r_{\text{out}}=1.255$,
corresponding to critical impact parameters $b_{\text{in}}=2.322$
and $b_{\text{out}}=3.515$, respectively. Our numerical findings
suggest that light rays with $n=3$ and $4$ typically orbit between
the two photon spheres, while for $n>4$, light rays tend to encircle
either the inner or outer photon sphere closely. The top-left panel
of FIG. \ref{fig:dp} illustrates the observed intensity $I_{\text{o}}$
as a function of $\beta$, displaying contributions from the direct
emission, lensing ring and photon ring. In contrast to the scenario
with a single photon sphere, the contributions to $I_{\text{o}}$
from the lensing and photon rings are comparable. Remarkably, nearly
bound light rays associated with the inner and outer photon spheres
generate two pronounced peaks in the intensity profile, located at
$\beta_{\text{in}}\simeq0.023$ and $\beta_{\text{out}}\simeq0.035$,
respectively. When compared to the single-photon sphere case, the
presence of light rays orbiting between these two photon spheres not
only broadens the width of the photon ring but also introduces a richer
structural complexity between the two distinct peaks. As depicted
in the top-right panel of FIG. \ref{fig:dp}, the presence of double
photon spheres substantially amplifies the intensity flux of the black
hole image, given the non-negligible contribution of the photon ring
to the total intensity.

The bottom panel of FIG. \ref{fig:dp} presents the corresponding
visibility amplitude $\left\vert V\left(u\right)\right\vert $ and
its individual components arising from the direct emission, lensing
ring and photon ring. Notably, while the direct emission and lensing
ring components share similarities with their counterparts in the
single-photon sphere case, the photon ring component displays marked
differences due to the presence of two peaks and the intricate internal
structure between them in the intensity profile of the photon ring.
At short baselines, the radial profiles of both the lensing and photon
rings remain unresolved. Due to their comparable contributions to
the intensity, their visibility components exhibit similarity. However,
as the baseline length increases, the internal structure between the
peaks of the photon ring becomes resolved, causing the contribution
from this internal structure to decay exponentially. As a result,
the photon ring component primarily depends on the two intensity peaks
at $\beta=\beta_{\text{in}}$ and $\beta=\beta_{\text{out}}$, which
can be well approximated by the double-thin-rings model mentioned
previously. Furthermore, the total visibility is dominated by the
photon ring component at long baselines. This dominance results in
beat signals manifesting in the visibility amplitude for $u\gtrsim350$.
Additionally, the bottom panel of FIG. \ref{fig:dp} presents the
time intervals between these beat signals, demonstrating good agreement
with $1/\left(\beta_{\text{out}}-\beta_{\text{in}}\right)\simeq83.33$.
Moreover, this concordance improves with increasing baseline length,
as the contribution from the internal structure diminishes for longer
baselines.

\section{Conclusions}

\label{sec:Conclusions}

In this paper, we have investigated the interferometric signatures
of scalarized RN black holes within the framework of the EMS model.
The EMS model accommodates the existence of scalarized black holes
endowed with double photon spheres, which can significantly impact
the observable features of black hole images. We began by discussing
the general framework of interferometric observations, focusing on
the complex visibility and its connection to the observed intensity.
Through the adoption of a simplified model that conceptualizes the
photon ring as one or two thin, uniform and circular rings on the
image plane, we derived analytical expressions for the complex visibility.
Subsequently, numerical simulations were executed to generate images
depicting black holes possessing either a single photon sphere or
double photon spheres, accompanied by an exploration of their corresponding
visibility signals.

Our findings confirm the characteristic damped oscillations of the
visibility amplitude for black holes with a single photon sphere,
with the photon ring dominating at long baselines. However, the presence
of double photon spheres in scalarized RN black holes introduces distinct
features, setting them apart from their single-photon-sphere counterparts.
A salient attribute of this phenomenon is the non-negligible contribution
of the photon ring to the observed brightness, manifesting as two
distinct peaks in the intensity profile. Furthermore, the interference
of light rays orbiting the two photon spheres generates beat signals
in the visibility amplitude, offering a unique feature potentially
detectable in future observations. The time intervals between these
beat signals can be accurately estimated using the double-thin-rings
model.

These results highlight the importance of interferometric studies
in probing the nature of compact objects and testing alternative gravity
theories. The distinctive features of black holes with multiple photon
spheres offer promising avenues for differentiating them from their
Kerr counterparts through high-resolution interferometric observations.
Further investigations, including the impact of realistic accretion
disk models and observational constraints, will contribute significantly
to the ongoing quest to unravel the fundamental properties of black
holes and gravity.
\begin{acknowledgments}
We are grateful to Guangzhou Guo and Tianshu Wu for useful discussions
and valuable comments. This work is supported in part by NSFC (Grant
No. 12275183, 12275184 and 11875196). 
\end{acknowledgments}

 \bibliographystyle{unsrturl}
\bibliography{ref}

\end{document}